\documentclass{article}

\usepackage{graphicx}
\usepackage{authblk}

\usepackage{tikz}
\usepackage{comment}
\usepackage{amsmath,amssymb} 
\usepackage{color}

\usepackage[accsupp]{axessibility}  

\author{Robert Turnbull}

\affil[]{Melbourne Data Analytics Platform}
\affil[ ]{University of Melbourne, Victoria 3052, Australia}
\affil[ ]{robert.turnbull@unimelb.edu.au }
\affil[ ]{ https://mdap.unimelb.edu.au}

\title{Cov3d: Detection of the presence and severity of COVID-19 from CT scans using 3D ResNets} 

\begin{document}
\pagestyle{headings}



%
%
%
\maketitle

\begin{abstract}
Deep learning has been used to assist in the analysis of medical imaging. One such use is the classification of Computed Tomography (CT) scans when detecting for COVID-19 in subjects. This paper presents Cov3d, a three dimensional convolutional neural network for detecting the presence and severity of COVID19 from chest CT scans. Trained on the COV19-CT-DB dataset with human expert annotations, it achieves a macro f1 score of 0.9476 on the validation set for the task of detecting the presence of COVID19. For the task of classifying the severity of COVID19, it achieves a macro f1 score of 0.7552. Both results improve on the baseline results of the `AI-enabled Medical Image Analysis Workshop and Covid-19 Diagnosis Competition' (MIA-COV19D) in 2022.

\end{abstract}

\section{Introduction}

To best care for patients, medical professionals need fast and accurate methods for detecting the presence and severity of COVID-19 in patients. Nucleic acid amplification tests (NAAT), such as real-time reverse transcription polymerase chain reaction (rRT-PCR), are recommended by the World Health Organisation (WHO) because they are highly specific and sensitive~\cite{who2021}. These kinds of tests have the disadvantage of taking a long time and not occurring at the point of need~\cite{peeling2022}. Medical imaging can be used to complement these diagnostic strategies~\cite{kollias2022ai}. Computed Tomography (CT) scans use x-rays to reconstruct cross-sectional images to produce a three dimensional representation of the internals of the body~\cite{seeram2018}. Thoracic radiologists have used chest CT scans to correctly diagnose patients with COVID-19 including cases where rRT-PCR gives a negative result~\cite{doi:10.1148/radiol.2020200343}. But this technique requires a human interpreter with sufficient knowledge and experience to provide reliable results. Advances in deep learning (DL) for computer vision offers the possibility of using artificial intelligence for the task of CT scan image analysis~\cite{kollias2018deep}. Early in the pandemic, attempts to use deep learning showed significant promise for accurate detection of COVID-19 ~\cite{harmon2020,kollias2020deep,kollias2020transparent}. To further this area of research, the `AI-enabled Medical Image Analysis Workshop and Covid-19 Diagnosis Competition' (MIA-COV19D) was created as part of the International Conference on Computer Vision (ICCV) in 2021~\cite{kollias2021mia}. This competition sought submissions to predict the presence of COVID-19 in a dataset of CT scan cross-sectional images. The winning submission produced a macro F1 score of 90.43 on the test partition of the dataset~\cite{Hou2021}. In 2022, the AI-enabled Medical Image Analysis Workshop issued a second competition with an enlarged dataset and new task which is to also predict the severity of COVID-19 in pa~\cite{kollias2022ai}. This article presents an approach to the tasks of this competition by using a three dimensional convolutional neural network called Cov3d.\footnote{The code for this project has been released under the Apache-2.0 license and is publicly available at https://github.com/rbturnbull/cov3d.}

\section{The COV19-CT-DB Database}

\begin{figure}
\centering
\includegraphics[width=\textwidth]{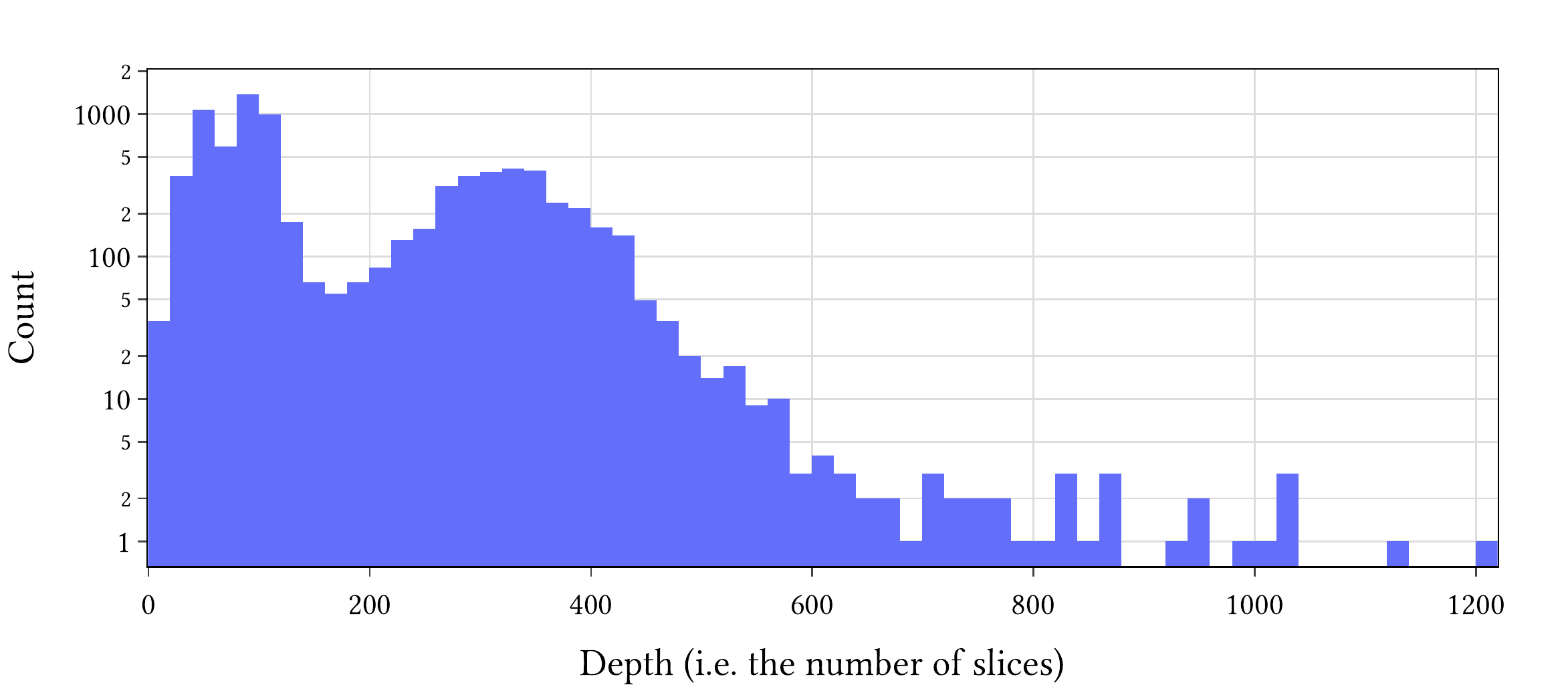}
\caption{A histogram of the number of slices NB. The vertical axis uses a logarithmic scale.}
\label{fig:depth-dist}
\end{figure}

The database (COV19-CT-DB) comprises more than 7,700 chest CT scans from more than 3,700 subjects and is annotated to record whether or not the subject had COVID-19~\cite{kollias2022ai}. Each CT scan contains 2D slices perpendicular to the long axis of the subject. The number of slices (hereafter referred to as the `depth') of the scan ranges from 1 to 1201. It has a bimodal distribution (see fig.~\ref{fig:depth-dist}) with more then 58\% of scans have fewer than 160 slices and almost 99\% having under 500 slices. The slices are provided as sequentially numbered JPEG files and typically have a resolution of 512$\times$512 pixels. It has been divided into training, validation and test partitions (table~\ref{table:partitions}).

\setlength{\tabcolsep}{4pt}
\begin{table}
\begin{center}
\caption{The number of CT scans in the partitions of the database}
\label{table:partitions}
\begin{tabular}{cccc}
\hline\noalign{\smallskip}
 COVID-19 & Training & Validation & Test\\
\noalign{\smallskip}
\hline
\noalign{\smallskip}
Positive & 882 & 215 & -- \\
Negative & 1,110 & 269 & -- \\
\hline
Total & 2,292 & 484  &  5,281 \\
\hline
\end{tabular}
\end{center}
\end{table}
\setlength{\tabcolsep}{1.4pt}

\setlength{\tabcolsep}{4pt}
\begin{table}
\begin{center}
\caption{The number of CT scans with severity annotations in the partitions of the database.}
\label{table:severity}
\begin{tabular}{ccccc}
\hline\noalign{\smallskip}
 Index & Severity & Training & Validation & Test\\
\noalign{\smallskip}
\hline
\noalign{\smallskip}
1 & Mild & 85 & 22 & -- \\
2 & Moderate & 62 & 10 & -- \\
3 & Severe & 85 & 22 & -- \\
4 & Critical & 26 & 5 & -- \\
\hline
& Total & 258 & 106 &  265 \\
\hline
\end{tabular}
\end{center}
\end{table}
\setlength{\tabcolsep}{1.4pt}

For a minority of the CT scans where COVID-19 was present, four experts have annotated the severity of disease in the patient with four categories: mild, moderate, severe and critical. These categories correspond to greater degrees of pulmonary parenchymal involvement. These annotations are indicated in CSV files that accompany the database. The number of scans with these annotations for the three partitions in the database are given in table~\ref{table:severity}.

\begin{figure}
\centering
\includegraphics[height=17cm]{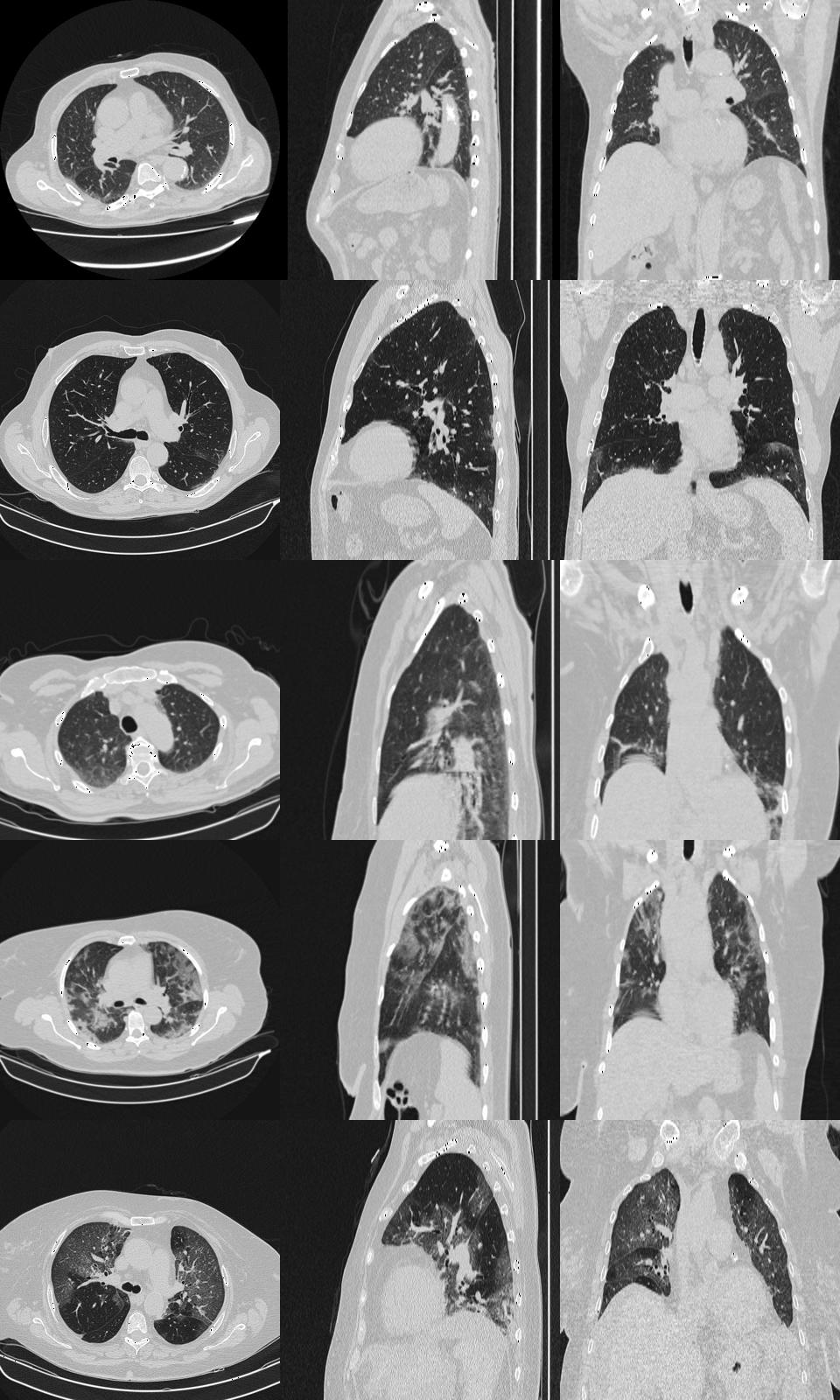}
\caption{Cross-sections from the five CT scans in the axial (left), sagittal (middle), and coronal (right) planes. The rows correspond to subjects without COVID19 (top) and then having COVID19 with severity categorizes as (in order): mild, moderate, severe and critical (bottom).}
\label{fig:cross-sections}
\end{figure}

\section{Methods}

To analyze this database, this paper discusses Cov3d, a classifier of CT scan images using three dimensional convolutional neural networks. Cov3d uses the deep learning framework PyTorch~\cite{NEURIPS2019_9015}. Cov3d also uses the fastai library which adds higher level components~\cite{fastai2}. It also is built using the FastApp framework for packaging deep learning models built with fastai and wrapping them in a command-line interface.\footnote{FastApp is available as an alpha release at: https://github.com/rbturnbull/fastapp/.}

\subsection{Preprocessing}

As seen in fig.~\ref{fig:depth-dist}, the number of slices in each CT scan is significantly varied. To regularize these so that multiple scans could be processed in batches and to reduce the size of the inputs to fit within the memory of the hardware, each scan went through a preprocessing step. First the 2D slices were reduced from a resolution of $512^2$ pixels to $128^2$, $256^2$ or $320^2$ pixels through bicubic interpolation. Then the slices were interpolated using 1D cubic interpolation along the axis perpendicular to the cross sections so that the depth was exactly half of the width/height of the processed scan. This gives three resolution sizes: small (64$\times$128$\times$128), medium (128$\times$256$\times$256) and large (160$\times$320$\times$320). These processed 3D images are saved as PyTorch tensor files and are loaded from disk during the training cycle. Only a single channel is stored for each image.

\subsection{Neural Network Model}

ResNet models have been tremendously popular for computer vision tasks on 2D images~\cite{resnet}. The residual `shortcut' connections between the layers address the problem of vanishing gradients which allows for models with a greater number of layers. Cov3d uses a neural network model analogous to a two dimensional ResNet-18 model but with 3D convolutional and pooling operations instead of their 2D equivalents. In particular, Cov3d uses the ResNet 3D 18 model included in the Torchvision library~\cite{Tran2018}. This model has been pre-trained on the Kinetics-400 dataset which classifies short video clips~\cite{Kinetics}. The time dimension in the pre-trained model was used for the depth dimension of the CT scans. Though the CT scans are quite different to video clip data in Kinetics-400, we can anticipate that the pre-trained network will have learned to identify shapes and patterns, particularly at the early layers of the network, which ideally will improve training through transfer learning (TL). 

A number of modifications to the pre-trained model were made before use. Since the pre-trained model used 3 channel inputs, the weights for the initial convolutional layer were summed across the channels so that single channel input could be used. Dropout~\cite{dropout} was added after each of the four main layers of ResNet model. The final linear layer to predict the 400 categories of the Kinetics dataset was replaced with two linear layers with a ReLU activation function between then and dropout. The size of the penultimate layer was varied as a hyperparameter and the size of the final layer was dependent on the loss function discussed below.

\subsection{Loss}

For detection of the presence of COVID-19, the penultimate layer connects to a single logit $x_i$ which models the log-odds that the subject has the disease, where $i$ refers to the index of the input. This can be converted to a probability $p_i$ using the sigmoid function:

\begin{equation}
    p_i = \frac{1}{1+e^{-x_i}}
\end{equation}

The loss for task 1 (i.e. detection of the presence of COVID-19) is denoted $\ell_{covid}$ and is given by the weighted binary cross entropy:

\begin{equation}
\ell_{covid}(p_i, y_i) = - w_i \left( y_i \log p_i + (1 - y_i)  \log (1 - p_i) \right)
\end{equation}

where $w_i$ is the weight of the input item and $y_i$ is the target probability. The weights are set to compensate for the different proportion of COVID-19 and non-COVID-19 scans in the training partition (see table~\ref{table:partitions}). 

Szegedy et al. introduced a regularization mechanism called label-smoothing which modifies the target probabilities to mitigate against the network becoming overconfident in the result~\cite{labelsmoothing}. This technique is employed here and thus the target probability for positive COVID19 scans is set to $1 - \epsilon_p$ and the target probability for negative COVID19 scans is $\epsilon_p$. The hyperparameter $\epsilon_p$ was set to either 0.0 or 0.1.


Task 2 requires inferring the severity of COVID-19 in patients base on the four categories. The dataset for this task is quite small (table~\ref{table:severity}) which makes training a model challenging. We can expand this dataset by regarding the COVID19 negative scans as belonging to a fifth category. Furthermore, the COVID19 positive scans which have not been added in the dataset can be regarded as belonging to a superset of those four severity categories. This can be modelled by connecting the penultimate layer to a vector of five dimensions ($z_{i,c}$) corresponding to the four severity categories ($c=1,2,3,4$) and an additional category for being COVID19 negative ($c=0$). The probability distribution over the categories ($s_{i,c}$) is given by the softmax function:

\begin{equation}
s_{i,c} = \frac{e^{z_{i,c}}}{\sum_{j=0}^4 e^{z_{i,j}}}
\end{equation}

The probability that the scan is merely COVID19 positive is given by summing over the four severity categories ($\sum_{c=1}^4s_{i,c}$). If the input is in one of the four annotated severity categories, or is classed as non-COVID19, then the loss is thus given with the cross entropy:

\begin{equation}
\ell_{severity}(s_{i,c}, y_{i}) = - w_i \sum_{c=0}^4 y_{i,c}\log(s_{i,c})
\end{equation}

where $w_i$ is the weight of the input item and $y_{i}$ is the target probability for class $c$. If the input is annotated as COVID19 positive but without a severity category then the loss is given by:

\begin{equation}
\ell_{severity}(s_{i,c}, y_{i}) = - w_i \log(\sum_{c=1}^4 s_{i,c})
\end{equation}

In this way, the second task to predict the severity can be trained using the entire database. Accuracy metrics for this task are restricted to the instances in the validation set which were annotated with the four severity categories.

As above, label smoothing was used as a regularization technique. However, unlike in Szegedy where the target probabilities were combined with a uniform distribution~\cite{labelsmoothing}, here the target probability for each class is reduced to $1 - \epsilon_s$ and the remaining probability of $\epsilon_s$ is divided between neighbouring categories whilst non-neighbouring categories remain with a target probability of zero. Non-COVID19 scans were considered to be neighbouring to `mild' severity scans.

The two loss functions discussed above could be used independently to train in separate models or the can be combined as a linear combination to train a single model to perform both tasks simultaneously:

\begin{equation}
\ell_{combined} = (1-\lambda) \ell_{covid} + \lambda \ell_{severity}
\end{equation}

where $0.0 \leq \lambda \leq 1.0$.



\subsection{Training Procedure}
The models were trained for 30, 40 or 50 epochs through the training dataset. The batch size was limited to two because of memory constraints. The Adam optimization method was used for updating the training parameters~\cite{adam}. The maximum learning rate was set to $10^{-4}$ and this and the momentum for the optimizer were scheduled according to the `1cycle' policy outlined by Smith~\cite{smith2018}. Scikit-learn was used to calculate the macro f1 scores on the validation set every epoch~\cite{scikit-learn}. The weights which yielded the highest macro f1 score on the validation dataset for the two tasks are saved for later inference.

\subsection{Regularization and Data Augmentation}

To mitigate against overfitting on the training dataset, there is the option to randomly reflect the input scans through the sagittal plane each training epoch. At inference, the reflection transformation is then applied to the input and the final probability predictions are taken from the mean. Weight decay of $10^{-5}$ was applied.

\section{Results}

The models were trained using NVIDIA Tesla V100-SXM2-32GB GPUs on the University of Melbourne's high-performance computing system Spartan. Results were logged using the `Weights and Biases' platform for experiment tracking~\cite{wandb}.

\setlength{\tabcolsep}{4pt}
\begin{table}
\begin{center}
\caption{Results of experiments. The best result for each task is highlighted in bold. In descending order of task 1 macro f1.}
\label{table:results}
\begin{tabular}{cccccccccc}
\hline\noalign{\smallskip}
ID & Depth & \shortstack{Width/ \\ Height} & Reflection & $\lambda$ & Epochs & $\epsilon_{p}$ & $\epsilon_{s}$ & \shortstack{Task 1 \\ macro f1} & \shortstack{Task 2 \\ macro f1} \\
\noalign{\smallskip}
\hline
\noalign{\smallskip}

1  & 160   & 320          & Yes        & 0.1       & 40     & 0.1                & 0.0                & \textbf{0.9476} {[}1{]}        & 0.4764                \\
2  & 160   & 320          & Yes        & 0.1       & 50     & 0.1                & 0.0                & 0.9412 {[}2{]}        & 0.4417                \\
3  & 160   & 320          & No         & 0.1       & 30     & 0.1                & 0.1                & 0.9394 {[}3{]}        & 0.4508                \\
4  & 160   & 320          & Yes        & 0.0       & 40     & 0.0                & 0.1                & 0.9372 {[}4{]}        & --                    \\
5  & 160   & 320          & Yes        & 0.5       & 30     & 0.1                & 0.1                & 0.9372                & 0.5952                \\
6  & 160   & 320          & Yes        & 0.9       & 50     & 0.1                & 0.0                & 0.9371                & 0.6369                \\
7  & 128   & 256          & Yes        & 0.1       & 40     & 0.1                & 0.0                & 0.9351                & 0.4424                \\
8  & 128   & 256          & Yes        & 0.1       & 50     & 0.1                & 0.0                & 0.9328                & 0.4812                \\
9  & 128   & 256          & Yes        & 0.1       & 40     & 0.0                & 0.0                & 0.9311                & 0.388                 \\
10 & 160   & 320          & Yes        & 0.1       & 30     & 0.1                & 0.1                & 0.9287                & 0.4153                \\
11 & 160   & 320          & Yes        & 0.9       & 40     & 0.1                & 0.0                & 0.9277                & 0.6623                \\
12 & 64    & 128          & Yes        & 0.1       & 40     & 0.1                & 0.1                & 0.9268                & 0.3988                \\
13 & 128   & 256          & No         & 0.5       & 30     & 0.1                & 0.1                & 0.9264                & 0.5909                \\
14 & 128   & 256          & Yes        & 0.9       & 40     & 0.1                & 0.0                & 0.9256                & \textbf{0.7546} {[}1{]}        \\
15 & 128   & 256          & No         & 0.1       & 30     & 0.1                & 0.1                & 0.9249                & 0.4416                \\
16 & 160   & 320          & Yes        & 0.9       & 40     & 0.1                & 0.1                & 0.9237                & 0.6766 {[}3{]}        \\
17 & 128   & 256          & Yes        & 0.0       & 40     & 0.0                & 0.1                & 0.9226                & --                    \\
18 & 64    & 128          & Yes        & 0.2       & 30     & 0.1                & 0.0                & 0.9222                & 0.4745                \\
19 & 64    & 128          & Yes        & 0.1       & 50     & 0.1                & 0.0                & 0.9204                & 0.4725                \\
20 & 64    & 128          & Yes        & 0.0       & 40     & 0.1                & 0.0                & 0.9183                & --                    \\
21 & 64    & 128          & Yes        & 0.1       & 40     & 0.1                & 0.0                & 0.9183                & 0.4376                \\
22 & 64    & 128          & Yes        & 0.0       & 40     & 0.0                & 0.1                & 0.9183                & --                    \\
23 & 128   & 256          & Yes        & 0.1       & 30     & 0.1                & 0.1                & 0.9167                & 0.4354                \\
24 & 64    & 128          & No         & 0.1       & 30     & 0.1                & 0.0                & 0.9093                & 0.3873                \\
25 & 64    & 128          & Yes        & 0.9       & 40     & 0.1                & 0.0                & 0.9062                & 0.6754 {[}4{]}        \\
26 & 64    & 128          & No         & 0.5       & 30     & 0.1                & 0.0                & 0.8973                & 0.6143                \\
27 & 128   & 256          & Yes        & 1.0       & 40     & 0.0                & 0.0                & --                    & 0.6712                \\
28 & 64    & 128          & Yes        & 1.0       & 40     & 0.0                & 0.0                & --                    & 0.6811 {[}2{]}       
              
\\

\hline
  
\end{tabular}
\end{center}
\end{table}
\setlength{\tabcolsep}{1.4pt}

The results of experiments with different hyperparameter settings are shown in table~\ref{table:results}. No one set of hyperparameters achieved the highest result in both tasks so two separate models are stored for inference on the two tasks. The highest macro f1 score for task 1 was 0.9476 which is significantly above the baseline of 0.77. This model used the highest resolution (160$\times$320$\times$320) and had a $\lambda$ value of 0.1 which weighs the loss heavily to $\ell_{covid}$. The highest macro f1 score for task 2 was 0.7552 which is above the baseline of 0.63. 

The submission for the test set includes the four highest performing models for each task (marked [1]--[4]) and also a basic ensemble of the four which averages the probability predictions between each model.

[RESULTS TO FOLLOW ON THE TEST SET ONCE THE INFORMATION IS RELEASED]

\section{Conclusion}

Cov3d is a three dimensional model for detecting the presence and severity of COVID19 in subjects from chest CT scans. It is based on a 3D ResNet pretrained on video data. The model was trained using a customized loss function to simultaneously predict the dual tasks in the `AI-enabled Medical Image Analysis Workshop and Covid-19 Diagnosis Competition'. The results for both tasks improve upon the baseline results for the challenge. Cov3d shows that deep learning can be used to interpret medical imaging such as CT scans and holds promise for complementing an array of other diagnostic methods to provide better care for patients.

\section{Acknowledgements}

This research was supported by The University of Melbourne’s Research Computing Services and the Petascale Campus. Sean Crosby and Naren Chinnam were instrumental in arranging the computational resources necessary. David Turnbull gave feedback on an earlier form of this paper which clarified the expression.

\clearpage
%
%
\bibliographystyle{splncs04}
\bibliography{references}
\end{document}